\documentclass[12pt]{iopart}

\usepackage{iopams}  
\usepackage[utf8]{inputenc}
\usepackage{aas_macros}
\usepackage{graphicx}
\usepackage{float}
\newcommand{\comm}[1]{}
\begin{document}

\title[Tachyon vs Quintessence Linear Theory]{Tachyonic vs Quintessence dark energy: linear perturbations and CMB data}

\author{Manvendra Pratap Rajvanshi$^1$, Avinash Singh$^2$, H.K. Jassal$^3$ \& J.S. Bagla$^4$}

\address{Dept. of Physical Sciences, IISER Mohali, Sector 81, SAS Nagar, Punjab (India) - 140306}
\ead{$^1$\mailto{manvendra@iisermohali.ac.in}, $^2$\mailto{avin.phy@gmail.com}, $^3$\mailto{hkjassal@iisermohali.ac.in}, $^4$\mailto{jasjeet@iisermohali.ac.in}}
\vspace{10pt}
\begin{indented}
\item[]April 2021
\end{indented}

\begin{abstract}
We use linear perturbation theory to study
perturbations in dynamical dark energy models.
We compare quintessence and tachyonic dark energy models with
identical background evolution. 
We write the corresponding equations for different models in a form
that makes it easier to see that the two models are very hard to
distinguish in the linear regime, especially for models with $(1 + w)
\ll 1$.
We use Cosmic Microwave Background data and parametric representations
for the two models to illustrate that they cannot be distinguished for
the same background evolution with existing observations.
Further, we constrain tachyonic models with the Planck data.
We do this analysis for exponential and inverse square potentials and
find that the intrinsic parameters of the potentials remain very
weakly constrained.
In particular, this is true in the regime allowed by low redshift
observations.  
\end{abstract}

\section{Introduction}

Observational evidence \cite{Riess_1998,Perlmutter_1999,1998ApJ...507...46S} of the
accelerated expansion of the Universe spurred the search for
theoretical models to explain the phenomenon.
In these models, the candidate that drives this acceleration is an
exotic, non-luminous, negative pressure medium, and 
it contributes approximately two-third of the energy budget of the
Universe at present.
This component called the dark energy
(DE) \cite{amendola_tsujikawa_2010,2006IJMPD..15.1753C,2012Ap&SS.342..155B} can be one of 
several theoretical possibilities including modifications to the
theory of gravity \cite{2016ARNPS..66...95J,2012Ap&SS.342..155B}.
The standard model of
cosmology ($\Lambda CDM$) represents dark energy by the cosmological
constant: this is equivalent to an effective density and pressure that
are constant in space-time.
All other models are dynamical in nature. They can have time-varying $w(z)$ and the observations sensitive to background expansion have been used by various authors to study these models \cite{2017JCAP...06..012T,Singh_2019,2020arXiv200600244B}. 
The existence of perturbations is a generic feature of dynamical dark
energy models and hence offers a probing ground for theories.
One of the most well-studied models of dark energy is
quintessence
\cite{Caldwell:1997ii,Ratra:1987rm,Caldwell:2005tm,Tsujikawa:2013fta}: 
the minimally coupled scalar field.
Another example is the tachyonic scalar field
model \cite{Padmanabhan:2002cp,Bagla:2002yn}.
These and other dark energy models have a scale-dependent
response to perturbations in the matter.
Linear theory of cosmological perturbations
\cite{10.1143/PTPS.78.1,MUKHANOV1992203,Ma:1995ey} is used for
studying the evolution of fluctuations in dark energy. 
These can be used to compute transfer functions, cosmic microwave
radiations anisotropies
\cite{Hu_Dodelson_doi:10.1146/annurev.astro.40.060401.093926}, and
other observables.  

Dark Energy perturbations have been studied in detail for quintessence
\cite{PhysRevD.78.123504,PhysRevD.81.083513,2009PhRvD..79l7301J,2012PhRvD..86d3528J,PhysRevD.64.103509}.
Perturbation theory employs a split between background and
perturbations over that background.
Comparisons of dynamical dark energy models with standard $\Lambda
CDM$ show deviations in expansion history.
Dark energy perturbations also induce differences in the power
spectrum, CMB anisotropies, and other observables.
For models allowing perturbations, the evolution is parameterized by a
combination of background evolution (or expansion history) and
characteristics of the model. 
This point needs careful consideration when comparing dark
energy models.
Their potentials often have tunable parameters that can be adjusted to
get the same expansion history so that any observation based on
background cosmology (e.g. supernova data) cannot be used to distinguish
these.
In that case, a question that can be asked is if perturbations-based
observations show any difference between such models? 
We address this question from the perspective of linear theory.
Further, we constrain tachyonic models with CMB data.
Tachyonic models have been constrained with low redshift observations
by Singh et. al \cite{Singh_2019,Singh_2020}.
Here we use data from Planck 2018 data in addition to other
observational datasets to constrain tachyonic models.\\ 

We present the results for quintessence and tachyonic models where we
compare the growth of perturbations in linear theory.
In the next two sections, we describe the formalism for writing
perturbation equations that demonstrate explicitly that the difference
between these two models diminish as we go towards $w=-1$.
Specifically, it depends on factor $(1+w)$, where $w$ is the equation
of state parameter.
We demonstrate that differences between the two classes of models are
suppressed by the factor $(1+w)$ and hence diminish rapidly as we
approach closer to the cosmological constant. 
We also relate our formalism with earlier work as well as the fluid
description.
In section 4, we present the numerical results.
In section 5, we use an approximate parametric representation for
quintessence and tachyonic models to use CMB data to see if these can
be distinguished by current data.
We find that if the respective potentials for tachyonic and
quintessence models are chosen such that the background evolution is
identical then it is not possible to distinguish these with present
observations.
Finally, in section 6, we constrain two tachyonic models with CMB
data. We summarize and comment on prospects in the last section. 
    
\section{Perturbation Theory}

We consider scalar metric perturbations in the Newtonian gauge with
the following form of the metric:
\begin{equation}
    ds^2 = (1+2\psi)dt^2 - (1-2\xi)a^2(dx^2 +dy^2 +dz^2)
    \label{eqn1}
\end{equation}
The problem of dark energy perturbations and its relevance for
observables has been studied by many authors \cite{2015MNRAS.452.2930M,Singh_2020,Bean:2003fb,PhysRevD.81.103513,PhysRevD.78.123504,PhysRevD.81.083513,2009PhRvD..79l7301J,2012PhRvD..86d3528J,PhysRevD.64.103509,2003MNRAS.346..987W}. 
Often, a fluid form for dark energy is assumed.
This fluid at the background level is characterized by the equation of
state parameter ($\bar{w}(z)$). 
For properly treating perturbations in fluid, one needs to know how
energy density perturbations as well as pressure perturbations
evolve. 
One can obtain dynamical equations for density perturbations and
velocity perturbation either directly from Lagrangian density or from
continuity equations (see Kodama \& Sasaki \cite{10.1143/PTPS.78.1},
Bean \& Dore \cite{Bean:2003fb} or Mukhanov \cite{MUKHANOV1992203}, Ma
\& Bertschinger \cite{Ma:1995ey}).
The set of equations derived in this manner is complete if additional
information is provided for term ${\delta p}/{\delta \rho}$.
In Newtonian gauge, ignoring anisotropic stress, we
have \cite{Ma:1995ey}: 
\begin{equation}
\dot{\delta} = -3H\left[\left( \frac{\delta p}{\delta \rho}\right) - \bar{w} \right]\delta + 3(1+\bar{w})\dot{\psi} - \frac{1+\bar{w}}{a}\theta
    \label{eq_fluid_pert1}
\end{equation}
\begin{equation}
    \dot{\theta} = -(1-3\bar{w})H\theta -\frac{\dot{\bar{w}}}{1+\bar{w}} + \frac{k^2 \psi}{a} + 
    \frac{k^2\delta}{(1+\bar{w})a}\left(\frac{\delta p}{\delta \rho}\right)
\end{equation}
where $\delta$ is the fluid energy density contrast, $\theta$ is defined as:
\begin{equation}
    \theta \bar{\rho}(1+\bar{w}) = - ik^j(\delta T)_j^0
    \label{eq_theta_def}
\end{equation}
For adiabatic perturbations in fluids, there exists a relation between
pressure perturbations and density perturbations.
Using this relation we can eliminate pressure perturbations and solve
for density perturbations.
In general, one has to solve for both perturbations.
In cases where density and pressure are effective quantities, e.g.,
scalar fields, the underlying system of equations has to be solved.  
A common approach is to quantify the variation of pressure
perturbations using a gauge-invariant quantity called the effective speed
of sound $c_s^2$.
In an arbitrary gauge, pressure perturbations are written as
\cite{Bean:2003fb,Batista:2013oca}:  
\begin{equation}
  (\delta p) = c_s^2(\delta \rho) + 3\dot{a}(1+\bar{w})(c_s^2-c_a^2)\bar{\rho}\frac{\theta}{k^2}
  \label{eq_delta_p}
\end{equation}
There are a few subtle points that need to be considered while using
this definition:
\begin{itemize}
\item
  In order to ensure gauge invariance, $c_s^2$ is defined in terms of
  ${\delta p}/{\delta \rho}$ in a frame comoving with fluid,
  i.e., frame in which $\theta$ is zero.
  Then from eq.(\ref{eq_delta_p}), $c_s^2$ is just ${\delta p}/{\delta
    \rho}$ but in the frame comoving with fluid. 
\item
  In general, there are entropy perturbations as well.
  The gauge invariant amplitude of entropy perturbation is
  \cite{10.1143/PTPS.78.1,Bean:2003fb}:  
  \begin{equation}
    \bar{w}\Gamma = (c_s^2-c_a^2)\delta
    \label{eq_entropy}
  \end{equation}
  where $c_a^2$ is a quantity determined by quantities related to
  evolution of the model background: 
  \begin{equation}
    c_a^2 = \bar{w} - \frac{\dot {\bar{w}}}{3H(1+\bar{w})}
  \end{equation}
  For adiabatic perturbations, $\Gamma$ vanishes, and
  $c_a = c_s$.   
\item
  Equation (\ref{eq_delta_p}) can be derived \cite{Bean:2003fb}
  starting from eq.(\ref{eq_entropy}). $\delta$ in general is not
  gauge invariant, but $\delta$ in the frame comoving with the fluid
  is a gauge invariant quantity. $\delta$ and $\theta$ can be combined
  to form a gauge invariant quantity: 
  \begin{equation}
    \delta_{rf}=\delta + 3\dot{a}(1+\bar{w})\frac{\theta}{k^2}
    \label{eq_gi_delta}
  \end{equation}
  In eq.(\ref{eq_entropy}), the left hand side is gauge invariant
  implying that combination on the right-hand side is  gauge
  invariant too.
  Now we define $c_s^2$ as ${\delta p}/{\delta
    \rho}$ in the rest frame of fluid meaning each term on
  the right-hand side is individually gauge-invariant in the context of this
  definition.
  Then in any frame we can substitute for rest frame $\delta$ using
  the quantity in eq.(\ref{eq_gi_delta}) in eq.(\ref{eq_entropy}) and
  then obtain eq.(\ref{eq_delta_p}).
  For multicomponent systems, there can be an additional entropy
  perturbation besides intrinsic entropy perturbations
  \cite{10.1143/PTPS.78.1}.
  This can be due to difference in dynamics (different $c_a^2$) or
  due to non-minimal coupling.
  In such cases, working in terms of field variables is simpler and less prone to ambiguities.  
\item
  For scalar fields \cite{Erickson:2001bq}, let $L(X,\phi)$ be the
  Lagrangian density, where $X = \frac{1}{2} \partial_\mu \phi
  \partial^\mu \phi$ is the kinetic term while $\phi$ is the field.
  Rest frame for field is defined as the one in which ($\delta \phi$)
  vanishes. 
  In an arbitrary frame:
  \begin{equation}
    (\delta p) = \frac{\partial p}{\partial X} (\delta X) +
    \frac{\partial p}{\partial \phi} (\delta \phi)
  \end{equation}
  with similar equation for ($\delta \rho$).
  In rest frame:
  \begin{equation}
    (\delta p) = \frac{\partial p}{\partial X} (\delta X)
  \end{equation}
  Combining equations for $(\delta p)$ and $(\delta \rho)$ in rest frame, we get
  \begin{equation}
    c_s^2 = \frac{(\delta p)}{(\delta \rho)} = \frac{p_{,X}}{\rho_{,X}} 
  \end{equation}
  where $p_{,X}$ is partial derivative wrt $X$. 
\end{itemize}
Earlier work \cite{Erickson:2001bq,Bean:2003fb,2003PhRvD..67j3509D,2003MNRAS.346..987W,PhysRevD.81.103513,2015MNRAS.452.2930M} along these lines has assumed some form for $c_s^2$ and
then constrained $c_s^2$ and other parameters.
These forms are assumed independent of $\bar{w}(z)$, thus the model is
described by two functions. 
One general result from these studies is that the effects of different
$c_s^2$ but same $\bar{w}(z)$ on observables are significant only in
cases where dark energy has some significant contribution (at least a
few percent) at time of recombination
\cite{Erickson:2001bq,PhysRevD.81.103513}.
But this itself means that $\bar{w}(z)$ should be of such a form that
dark energy has a significant contribution at early times.   
For scalar fields, given the form for Lagrangian density, there is no
need of using any ad-hoc approximate form for $c_s^2$.
Equations for systems with scalar field perturbations can be written
entirely in terms of field perturbations (gauge-invariant) and
perturbations in other constituents.
But studies with effective parametrization of $c_s^2$ are useful
because they provide a general framework to compare different type of
Lagrangian densities.
Different Lagrangian densities may have different effective speeds of
sound.
For example, canonical scalar field Lagrangian of form: 
\begin{equation}
    L = X - V(\phi)
\end{equation}
always have $c_s^2 = 1$, while k-essence ones with the form:
\begin{equation}
    L = V(\phi)F(X)
\end{equation}
can have a time-dependent $c_s^2$.

In this work, we consider the question whether two different scalar
field Lagrangians (tachyonic and quintessence) with the same background
evolution can be distinguished at the level of linear perturbations. 
Instead of working with an assumed form of $c_s^2$ and using fluid
equations, we directly work with scalar fields and their
perturbations. 
Our model space is limited as we choose two specific Lagrangians, but our
calculations are concrete with few assumptions. 
Our choice of formalism is well motivated by the question: all
observations sensitive to background cosmology (only) give us a
certain evolution of background quantities, then that can be explained
by both quintessence and tachyonic models with corresponding
reconstructed potentials.
We explore if these models can be distinguished by observations sensitive to linear
perturbations? 

\section{Basic equations for scalar fields with effective fluid approach}

In this section, we derive equations for quintessence and tachyonic
fields.
For establishing correspondence between field description and
the effective fluid description we define a new perturbation quantity:
$u$ which is the deviation in the equation of state from background
homogeneous fluid.
The fluid description we employ is slightly different from the
standard approach but is useful in highlighting differences in
quintessence and tachyonic models.
We also give relations between standard fluid variables and the
variables used here.

Let $\Phi$ be the field for a scalar field representing dark
energy.
Then its stress energy tensor can be written as: 
\begin{equation}
  T_{\mu\nu} = (\rho + P)v_\mu v_\nu - P g_{\mu\nu}
  \label{eqn2}
\end{equation}
where 
\begin{equation}
  v_\nu = \frac{\partial_\nu \Phi}{\sqrt{\partial^\alpha \Phi \partial_\alpha \Phi }}
\end{equation}
We define first order quantities, density contrast and the
corresponding variation in the equation of state parameter. 
\begin{equation}
  \rho = \bar{\rho}(1+\delta) \qquad   W = \bar{w}(1+u)
\end{equation}
where variables with a bar are background quantities dependent on time
only, while the first order variations ($\delta$ \& $u$) can vary in
space-time.  
We also define
\begin{equation}
  \omega = 1+\bar{w}
\end{equation}
Effective pressure ($P$) for a scalar field theory (with
identification of $P$ as per equation eq.(\ref{eqn2})) is the
Lagrangian ($L_\Phi$) of field while the effective density $\rho$ is:  
\begin{equation}
  \rho =2 g^{\mu \nu}\frac{\partial L_\Phi}{\partial g^{\mu \nu}} - L_\Phi
\end{equation}
Writing the field as the sum of background and perturbation:
\begin{equation}
  \Phi = \phi + (\delta \phi) 
\end{equation}
and substituting it in equation eq.(\ref{eqn2}), retaining only the
first order terms, we get the first order stress energy tensor using
metric eq.(\ref{eqn1}): 
\begin{eqnarray}
  T^0_0 &=& \bar{\rho}\delta \nonumber \\
  T^i_j &=& \bar{\rho}(u+\delta)(1-\omega)\quad\quad for\quad i=j
  \nonumber \\
  T^0_j &=&
  \frac{\bar{\rho}\omega}{\dot\phi}\frac{\partial(\delta\phi)}{\partial
    x^j} 
\end{eqnarray}
Off-diagonal spatial components of stress-energy tensor of both dark
matter and field vanish at this order, hence the two metric potentials
can be taken to be equal.
We choose to work with $\psi$. 
The first order Einstein equation 
\begin{equation}
  G^1_1 = 8\pi G T^1_1
\end{equation}
can be used to obtain
\begin{equation}
  \ddot{\psi}+4\frac{\dot a}{a}\dot\psi+\psi\left[
    \frac{2\ddot{a}}{a}+\frac{\dot a^2}{a^2}\right] = -4\pi
  G\bar{\rho}(u+\delta)(-\bar{w}) 
  \label{eqn12}
\end{equation}
We obtain the dynamical equations for $\delta$ and $u$ by requiring
that the four divergence of stress energy tensor vanishes, i.e. 
\begin{equation}
    T^\mu_{\nu;_\mu}=0
\end{equation}
\begin{equation}
  \dot\delta =  3u(1-\omega)\frac{\dot a}{a} + \omega\left[ 3\dot\psi
    + \frac{\nabla^2(\delta\phi)}{a^2\dot\phi}\right]  
    \label{eqn14}
\end{equation}
Making use of the following off-diagonal Einstein equation:
\begin{equation}
  \frac{\dot a }{a}\frac{\partial\psi}{\partial
    x^j}+\frac{\partial\dot\psi}{\partial x^j} = \frac{4\pi
    G\bar{\rho}\omega}{\dot\phi}\frac{\partial(\delta\phi)}{\partial
    x^j} + 4\pi G\quad ^{dm}T^0_j 
\end{equation}
with dark matter stress energy contribution as
\begin{equation}
  ^{dm}T^0_j = -a^2\bar{\rho}_{dm}\frac{\partial U}{\partial x^j}
\end{equation}
where $U$ is dark matter velocity potential, we rewrite equation
(\ref{eqn14}) as: 
\begin{equation}
  \dot\delta =  3u\frac{\dot a}{a}(1-\omega)+3\dot\psi\omega +
  \frac{1}{4\pi G \bar{\rho}a^2}\nabla^2\left[ \frac{\dot a }{a}\psi +
    \dot\psi\right] + \frac{\bar{\rho}_{dm}}{\bar{\rho}}\nabla^2 U 
  \label{eqn17}
\end{equation}
We also get a constraint equation for $u$
\begin{equation}
   \fl    (-1+\omega)\frac{\partial u}{\partial x^j} =
   (1-\omega)\frac{\partial \delta}{\partial x^j} +
   \frac{\omega}{\dot\phi}\frac{\partial(\delta\phi)}{\partial
     x^j}\left[3\frac{\dot a}{a} + \frac{\dot{\bar{\rho}}}{\bar{\rho}}
     +\frac{\dot\omega}{\omega}  \right] +
   \frac{\omega}{\dot\phi}\left[
     -\frac{\ddot{\phi}}{\dot\phi}\frac{\partial(\delta\phi)}{\partial
       x^j} + \frac{\partial(\dot{\delta\phi})}{\partial x^j} \right]
   -\omega\frac{\partial\psi}{\partial x^j}  
\label{eqn18}
\end{equation}
We observe that equations (\ref{eqn12}) and (\ref{eqn17}) do not have
explicit dependence on particular details of scalar field (whether it
is quintessence or tachyonic), but equation \ref{eqn18} does have such
a dependence.
Therefore any differences between models will arise from this
equation. 
We rewrite these equations in a less ``field-specific" form and find
that the equations in one of the theories have more terms.  
For tachyonic field, equation (\ref{eqn18}) can be written as:
\begin{eqnarray}
    \frac{(-1+\omega)}{2}\frac{\partial u}{\partial x^j} & = &
    (1-\omega)\frac{\partial\delta}{\partial x^j} +
    \left[3(1-\omega)\frac{\dot a}{a} +\frac{\dot\omega}{2\omega}
      \right]\nonumber \\ 
    && \left[\frac{1}{4\pi G\bar{\rho}}\left(\frac{\dot
        a}{a}\frac{\partial \psi}{\partial x^j} +
      \frac{\partial\dot\psi}{\partial x^j}  \right)    +
      a^2\frac{\bar{\rho}_{dm}}{\bar{\rho}}\frac{\partial U}{\partial
        x^j}  \right]\label{eqn19} 
\end{eqnarray}
While quintessence has extra terms in addition to those present in
equation (\ref{eqn19}): 
\begin{eqnarray}
    \frac{(-1+\omega)}{2}\frac{\partial u}{\partial x^j} &=&
    (1-\omega)\frac{\partial\delta}{\partial x^j} +
    \left[3(1-\omega)\frac{\dot a}{a} +\frac{\dot\omega}{2\omega}
      \right] \nonumber\\ 
    && \left[\frac{1}{4\pi G\bar{\rho}}\left(\frac{\dot
        a}{a}\frac{\partial \psi}{\partial x^j} +
      \frac{\partial\dot\psi}{\partial x^j}  \right)    +
      a^2\frac{\bar{\rho}_{dm}}{\bar{\rho}}\frac{\partial U}{\partial
        x^j}  \right] \nonumber\\ 
    && +\large\omega \left[\frac{3\dot a}{8 \pi G
        \bar{\rho}a}\frac{\partial}{\partial x^j}\left(\frac{\dot
        a}{a}\psi +\dot\psi  \right) + \frac{3\bar{\rho}_{dm} \dot a
        a}{2\bar{\rho}}\frac{\partial U}{\partial x^j} +
      \frac{1}{2}\frac{\partial\delta}{\partial x^j}\right] \label{eqn20}
\end{eqnarray}
Observing the third line in the above equation and comparing it with
the equation for tachyonic counterpart (\ref{eqn19}), we find that the
difference between two models is encoded in the terms multiplied by
$\omega=(1+\bar{w})$.
For the models constrained by observations, this number is small, much
smaller than unity.
Effectively this makes the differences between two models a second
order term. 

We relate $u$ to familiar quantities:
\begin{equation}
    \frac{(u+\delta)\bar{w}}{\delta} = \frac{(\delta p)}{(\delta \rho)}
\end{equation}
The effective ``velocity" perturbation (coming from
\ref{eq_theta_def}) for scalar field is: 
\begin{equation}
    \theta = \frac{k^2 (\delta \phi)}{a\dot{\phi}}
\end{equation}
and the effective speed of sound is:
\begin{equation}
    c_s^2 = \frac{(u+\delta)\bar{w} +
      3\frac{\dot{a}}{a}(1+\bar{w})c_a^2 \frac{(\delta
        \phi)}{\dot{\phi}}} {\delta +
      3\frac{\dot{a}}{a}(1+\bar{w})\frac{(\delta \phi)}{\dot{\phi}}} 
\end{equation}
As stated earlier, we do not need to incorporate an effective $c_s^2$
while working with fields because we have an analytical expression
that can be evaluated.
But for comparison of models, we derive approximate
effective $c_s^2$ for tachyonic field.
Please note that $c_s^2$ for quintessence is unity.

For tachyonic field the Lagrangian density is:
\begin{equation}
    L(X,\Phi) = -V(\Phi)\sqrt{1-2X} 
\end{equation}
In the comoving frame of a scalar field:
\begin{equation}
  c_s^2 = \frac{p_{,X}}{\rho_{,X}} = \frac{L_{,X}}{L_{,X}+2L_{,XX}X}
\end{equation}
In case of tachyonic field:
\begin{equation}
  c_s^2 = (1-2X) = -\bar{w}-(1+\bar{w})(\delta g^{00})_{rf} \approx-\bar{w}
  \label{cs2eqn}
\end{equation}
In linear theory approximation $c_s^2$ is just $-\bar{w}$ as
$(1+\bar{w})\ll 1$ for models allowed by observations and the second
term in eq.(\ref{cs2eqn}) is effectively of second order.
Most of the comparisons of $c_s^2$ in literature are between very 
different values of $c_s^2$ like between 1, 0.1, 0.01,0, etc.
While we see that for models allowed by background observations,
tachyonic $c_s^2$ is not very different from quintessence value of $1$.

In the following sections, we study the differences in the two models in
linear theory using field perturbations.
Note that one can either directly use equations derived from field
perturbations or the fluid perturbations ($u$ and $\delta$) equations
derived in this section as these are equivalent approaches.  

\section{Results for field-based comparisons}

We divide our discussion here into 2 subsections. In first we show
comparisons for quantities, that influence observables, like  metric
potential ($\psi$) and its derivative ($\dot{\psi}$).
In the second
subsection, we show differences in dark energy perturbations. 

\subsection{Influence on observables}

All observables are affected by metric coefficients.
The influence of these coefficients on dark matter linear growth rate is
used in calculating observational effects like matter clustering,
$\sigma_8$, growth index, etc.
Rate of change of potential ($\dot{\psi}$) affects CMB photons and
causes observable effects like
ISW \cite{2012PhRvD..86d3528J,Sachs_Wolfe_1967ApJ...147...73S}.  

We present $\psi$ and $\dot{\psi}$ for the following background
evolutions (characterized by $\bar{w}(a)$): 
\begin{itemize}
\item
  Constant $\bar{w}(z)$ for values: - 0.5 and -0.975 
\item
  Chevallier-Polarski-Linder (CPL)
  paramterization \cite{2001IJMPD..10..213C,2003PhRvL..90i1301L} 
  \begin{equation}
    \bar{w}= w_0+w_a(1-\frac{a}{a_0})
    \label{eqn:cpl}
  \end{equation}
  $\bar{w}(z)$  with parameters: $w_0 = -0.9$ and $w_a = -0.099$ 
\end{itemize}
Since differences in growth rate of perturbations with scale has been
seen mainly at very large scales
\cite{PhysRevD.78.123504,2018JCAP...06..018P,Rajvanshi_2020} we
present results for length-scales: $2000$~Mpc and $10000$~Mpc. 
The differences between two models peak approximately around
$10000$~Mpc length-scales.
At small scales, the growth of perturbations is suppressed, and at very
large scales the growth rate is independent of the speed of sound.
It is only in the transition region that we can expect to capture some
differences between models with the same expansion history but a
different $c_s^2$.  

\begin{figure}[hbt!]
  \centering
  \includegraphics[width=1.1\textwidth]{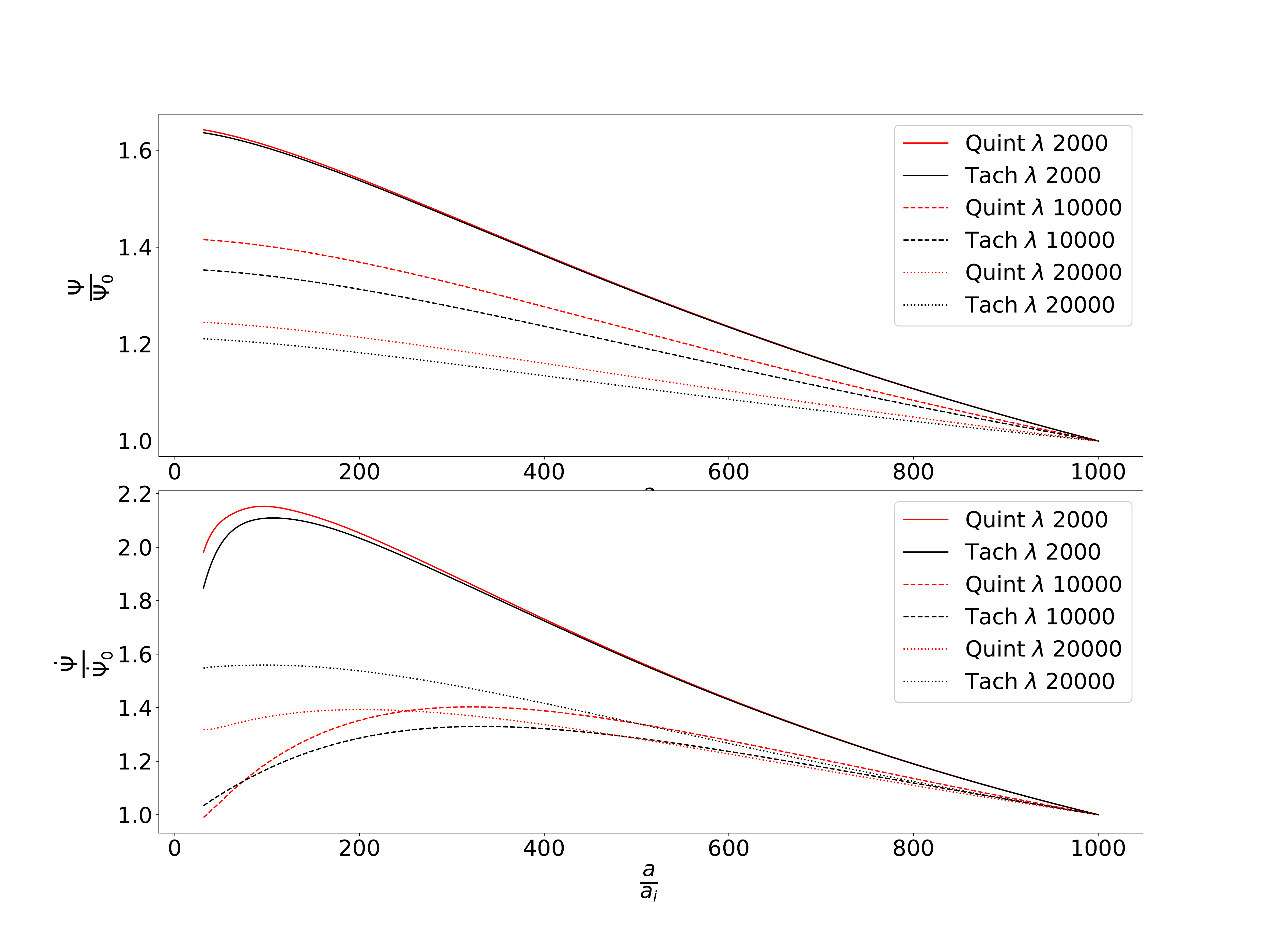}
  \caption{This figure shows the potential $\psi$ and its time derivatives for
    $\bar{w}=-0.5$. The potentials have been normalized by
    their present day value. The difference in different models is
    higher for $\lambda\sim 10k$.}
  \label{fig:psi_mp5}
\end{figure}

\begin{figure}[hbt!]
  \centering
  \includegraphics[width=1.1\textwidth]{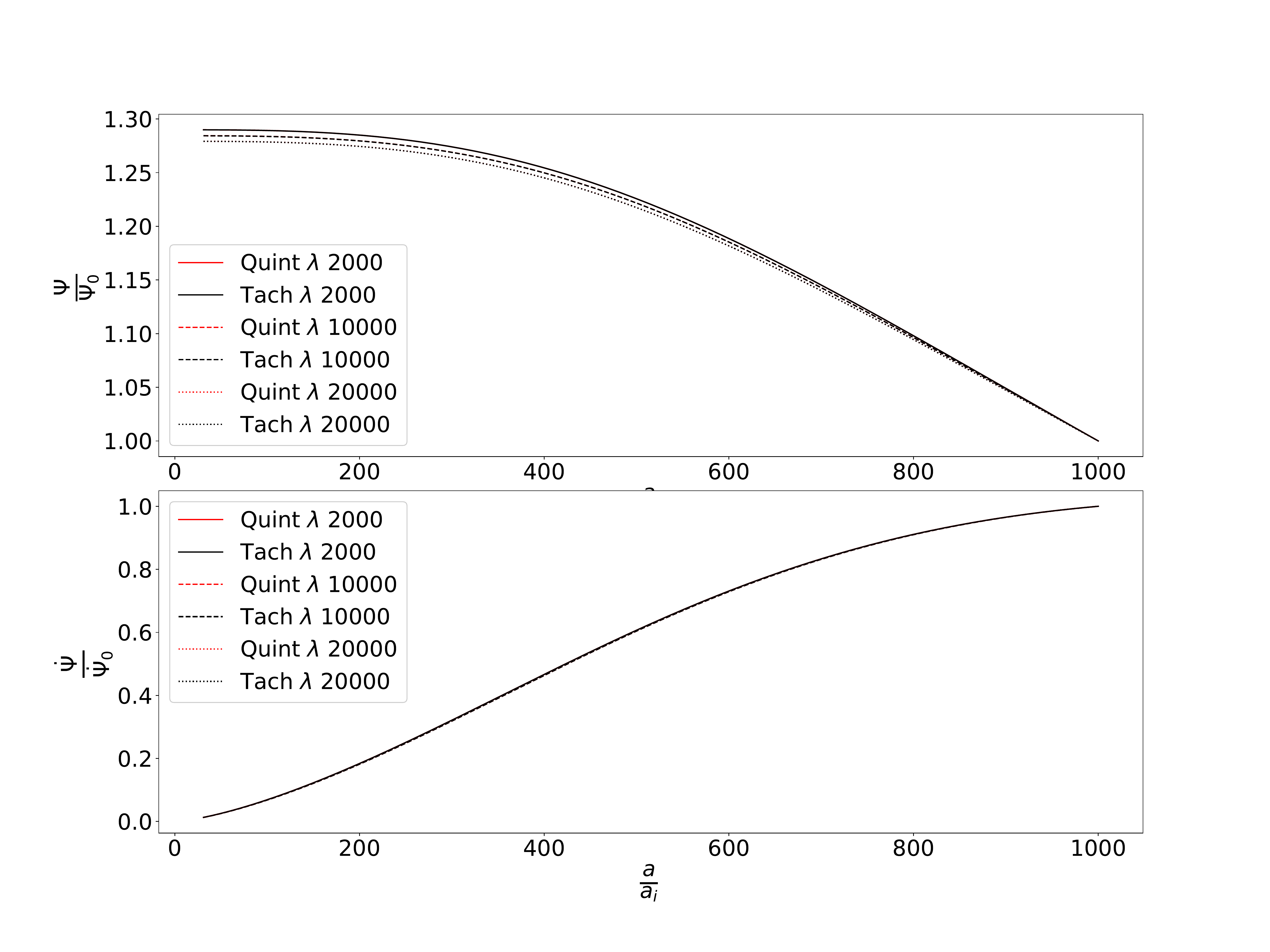}
  \caption{The plots show the potential $\psi$ and its time
    derivatives for $\bar{w}=-0.975$ case. Clearly, the  relative
    differences are much smaller than  in the case of $w=-0.5$.}
  \label{fig:psi_mp975}
\end{figure}

\begin{figure}[hbt!]
  \centering
  \includegraphics[width=1.1\textwidth]{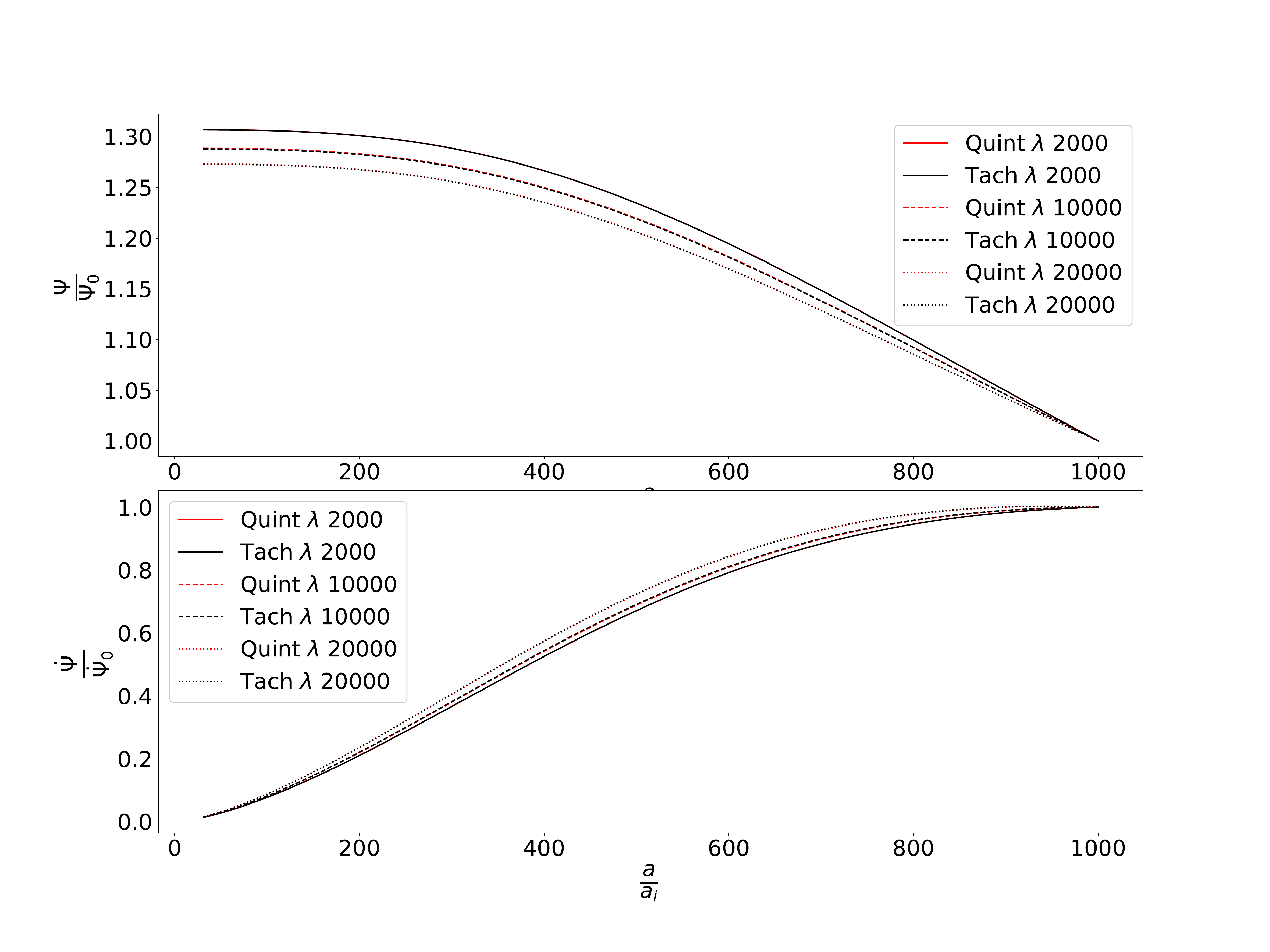}
  \caption{Here potential $\psi$ and its time derivatives are
    plotted  for the  CPL model. Differences are much smaller than
    $w=-0.5$ and slightly stronger than $w=-0.975$ case. This is
    because in comparison to $w=-0.975$, the background evolution
    for  the  CPL model  has a larger deviation from $w=-1$.} 
  \label{fig:psi_cpltest}
\end{figure}

We show $\psi$ and $\dot{\psi}$ in figures
\ref{fig:psi_mp5},\ref{fig:psi_mp975} and \ref{fig:psi_cpltest}.
In the notation used to annotate the curves, we use `quin' for
quintessence models and `tach' for the tachyon models. Also, we use red color for quintessence and black for the tachyonic field. The length scales
are mentioned alongside. 
We find that tachyonic and quintessence models for  $\bar{w}=-0.975$
and CPL cases are almost indistinguishable with differences of the order 
$0.01\%$ in most cases.
Corresponding differences for $\bar{w}=-0.5$ are more significant. 
These differences grow in a monotonic and continuous manner as we
moves away from $w = -1$.
We plot one extreme case $\bar{w}=-0.5$ that is observationally ruled
out but gives an indication of the order of differences between the two
classes of models.
CPL and $\bar{w}=-0.975$ are observationally
allowed \cite{2016A&A...594A..13P,2017JCAP...06..012T} but differences
between the models are extremely small at all scales. For $w=-0.5$, differences in potentials and its time derivatives can be of the order $10\%$. $w=-0.975$ case shows negligible differences of the order $0.01\%$ while CPL case has differences around $0.1\%$. 

\subsection{Scalar fields}

\begin{figure}[hbt!]
  \centering
  \includegraphics[width=1.2\textwidth]{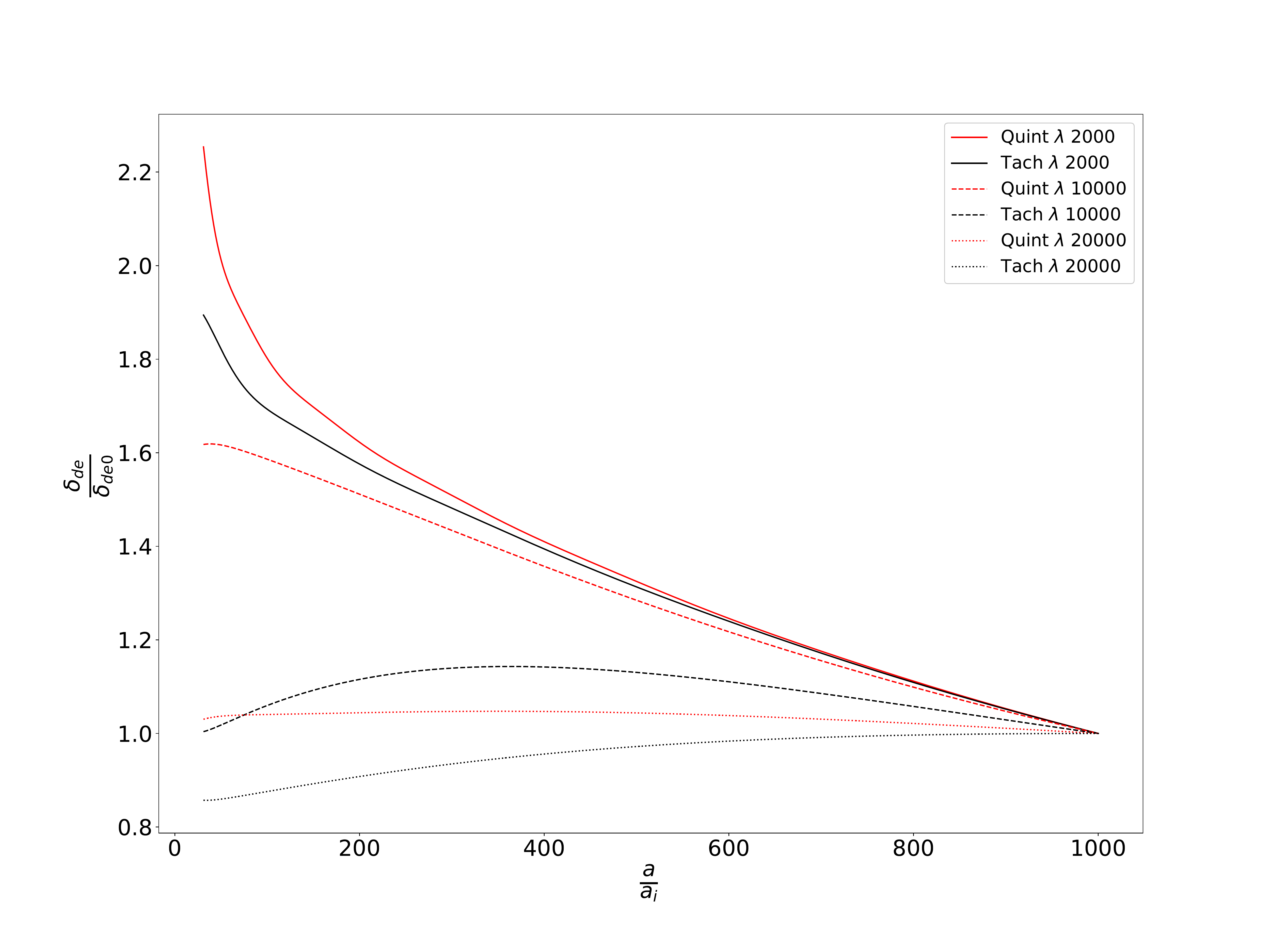}
  \caption{Dark energy (DE) density contrast for $\bar{w}=-0.5$
    case. It shows growth of DE normalized by present value.} 
  \label{fig:dc_de_mp5}
\end{figure}

\begin{figure}[hbt!]
  \centering
  \includegraphics[width=1.2\textwidth]{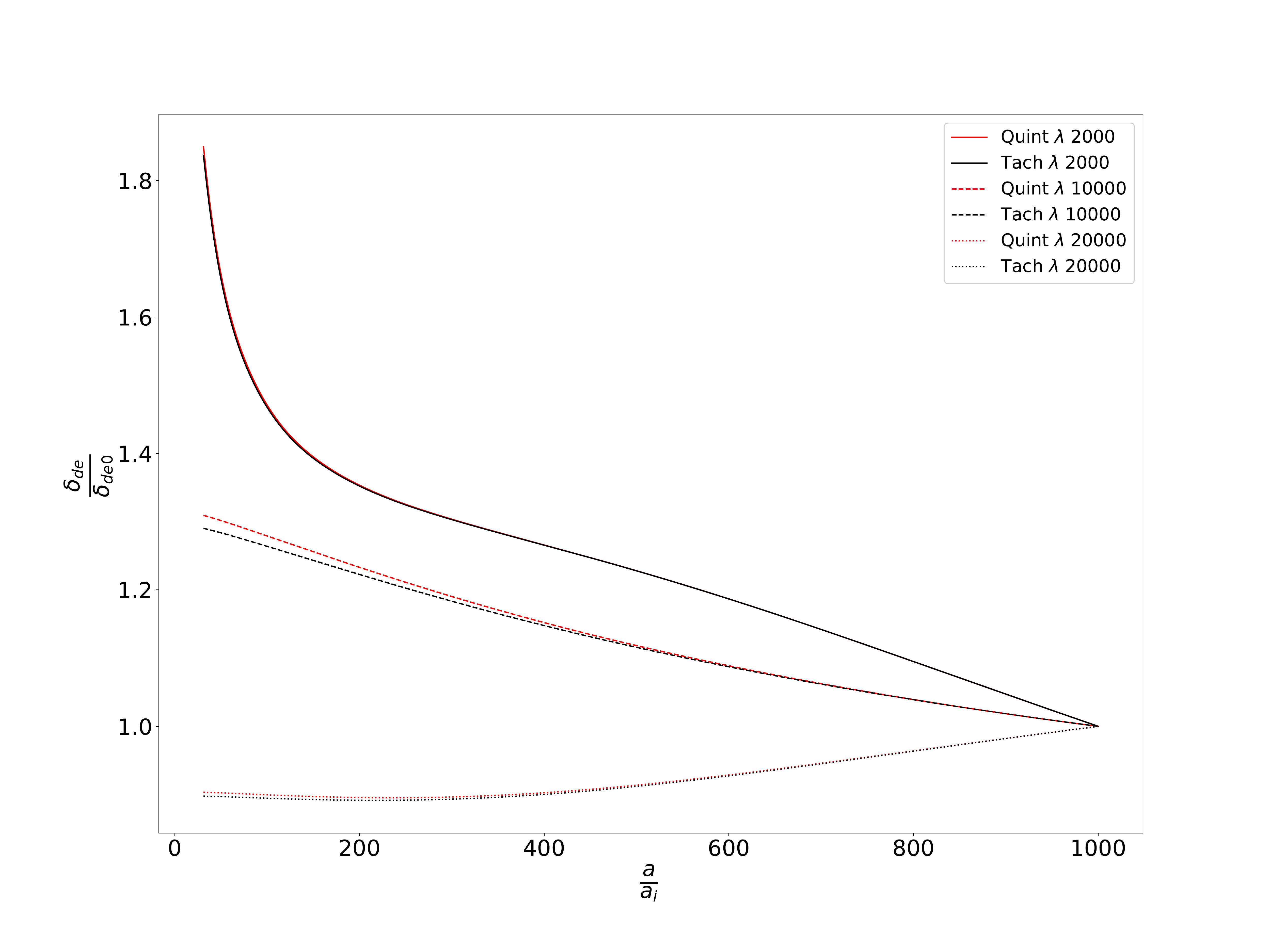}
  \caption{Dark energy (DE) density contrast for $\bar{w}=-0.975$
    case. Relative differences between quintessence and tachyonic
    models are small in comparison with $w=-0.5$ case.} 
  \label{fig:dc_de_mp975}
\end{figure}

\begin{figure}[hbt!]
  \centering
  \includegraphics[width=1\textwidth]{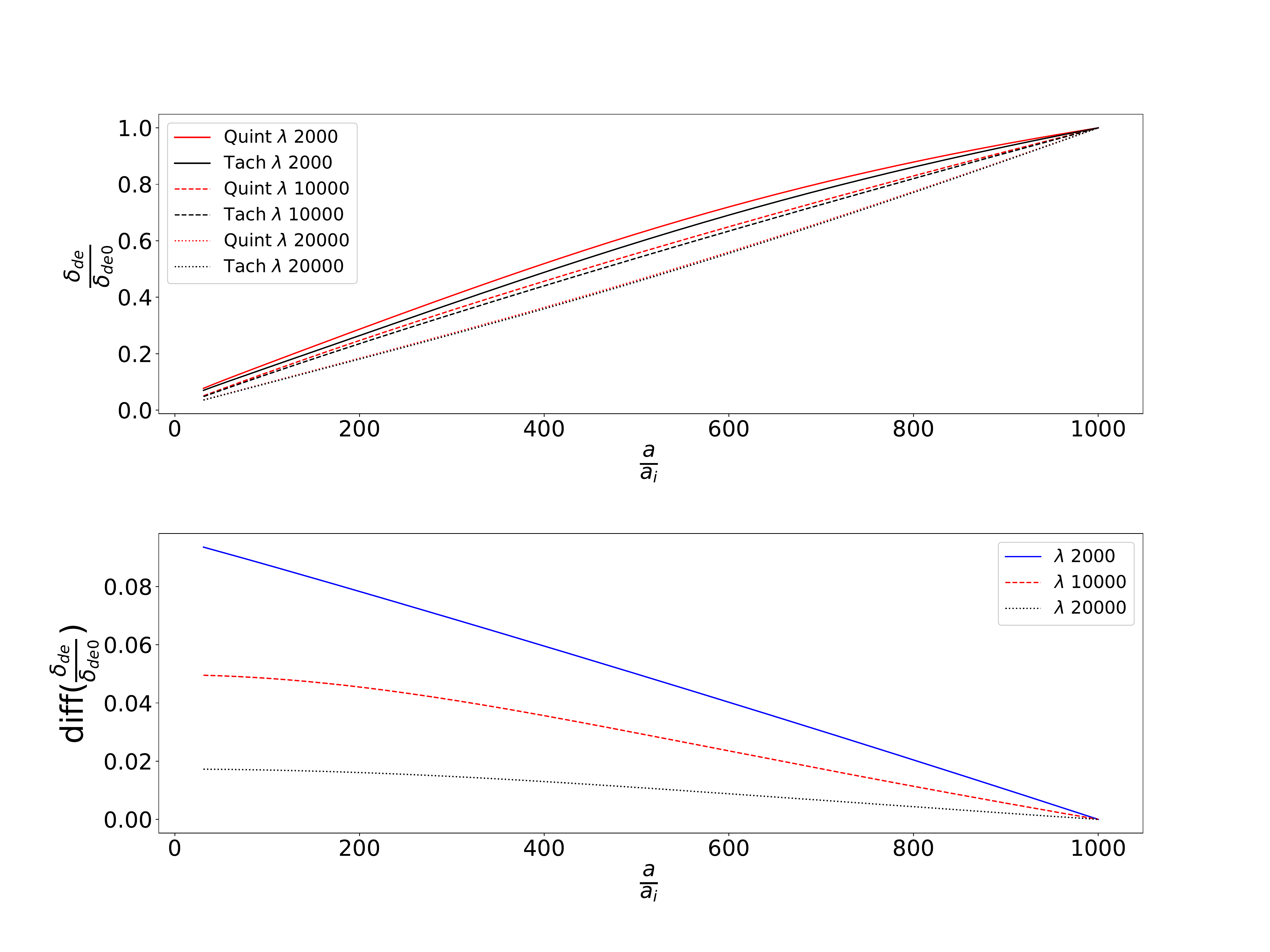}
  \caption{Dark energy (DE) density contrast for the CPL case. Curves
    are neatly clustered here by field type. This suggests that for
    this particular background evolution, tachyonic field and
    quintessence field evolve similarly for a particular
    lengthscale.} 
    \label{fig:dc_de_cpltest}
\end{figure}

While dark energy perturbations show more differences (figures \ref{fig:dc_de_mp5}, \ref{fig:dc_de_mp975} and
\ref{fig:dc_de_cpltest}) than
potentials, their effects on observables are not very
significant as shown in the previous subsection.
Fluctuations are stronger for cases that are significantly removed from
$w=-1$.
Since dark energy perturbations are not directly observable, 
the significance of fluctuations can only be evaluated through
observables as in the last subsection.
We have shown comparisons for dark energy perturbations in figures
\ref{fig:dc_de_mp5}, \ref{fig:dc_de_mp975} and
\ref{fig:dc_de_cpltest}.
Although there are visible differences (between tachyon and
quintessence cases) in the evolution of DE perturbations but these
differences remain insignificant because the amplitude of
perturbations is very small.  

\section{Constraining models with CMB anisotropy data}

There are two popular public codes available for CMB anisotropy
calculations: CAMB \cite{2000ApJ...538..473L} and CLASS
\cite{2011arXiv1104.2932L,2011JCAP...07..034B,2011arXiv1104.2934L}.
Both have support for implementing fluid models with effective
$c_s^2$.
Here we use CLASS to calculate CMB anisotropy power spectra
for effective $c_s^2$ corresponding to tachyon models and quintessence
models.
This requires some minor tweaks in the default CLASS code as
the standard version does not have time-dependent $c_s^2$.
We modified the code to allow for a time-varying form of $c_s^2$ for
tachyon models.
There are various ways tachyonic models can be included in
CLASS.
We can write effective potentials for the tachyon field in terms of a
chosen background DE (particular $w(a)$), or we can have an effective
fluid description with $c_s^2$ as derived in eq.(\ref{cs2eqn}).
While the former is a more apt and clean approach, the latter is easier to
implement and is expected to give same results for $(1+\bar{w})\ll 1$,
which anyway is the region already constrained by background cosmology
probes.
In cases where one has well-motivated forms for potentials, these
tachyonic models can be implemented in CLASS with some more effort. 
We do this in the next section where we constrain tachyon models for
two well-studied potentials. 

\begin{figure}[hbt!]
    \centering
    \includegraphics[width=1.2\textwidth]{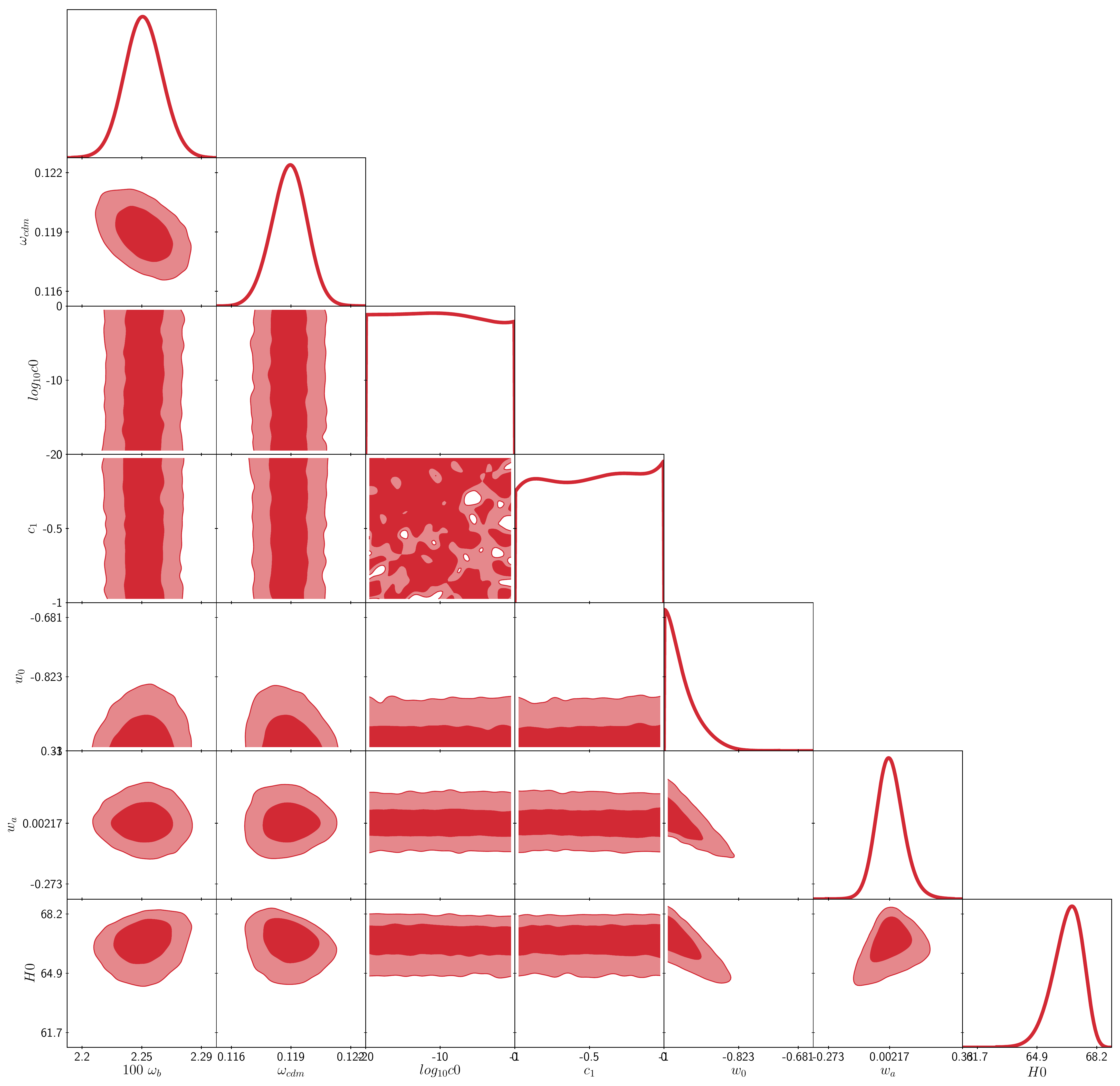}
    \caption{Triangle plot: While background cosmological parameters like density parameters, $H_0$, $w_0$,etc. are well constrained, perturbation related parameters like $c_s^2$ remain unconstrained. }
    \label{fig:traingle}
\end{figure}

\begin{figure}[hbt!]
    \centering
    \includegraphics[width=1\textwidth]{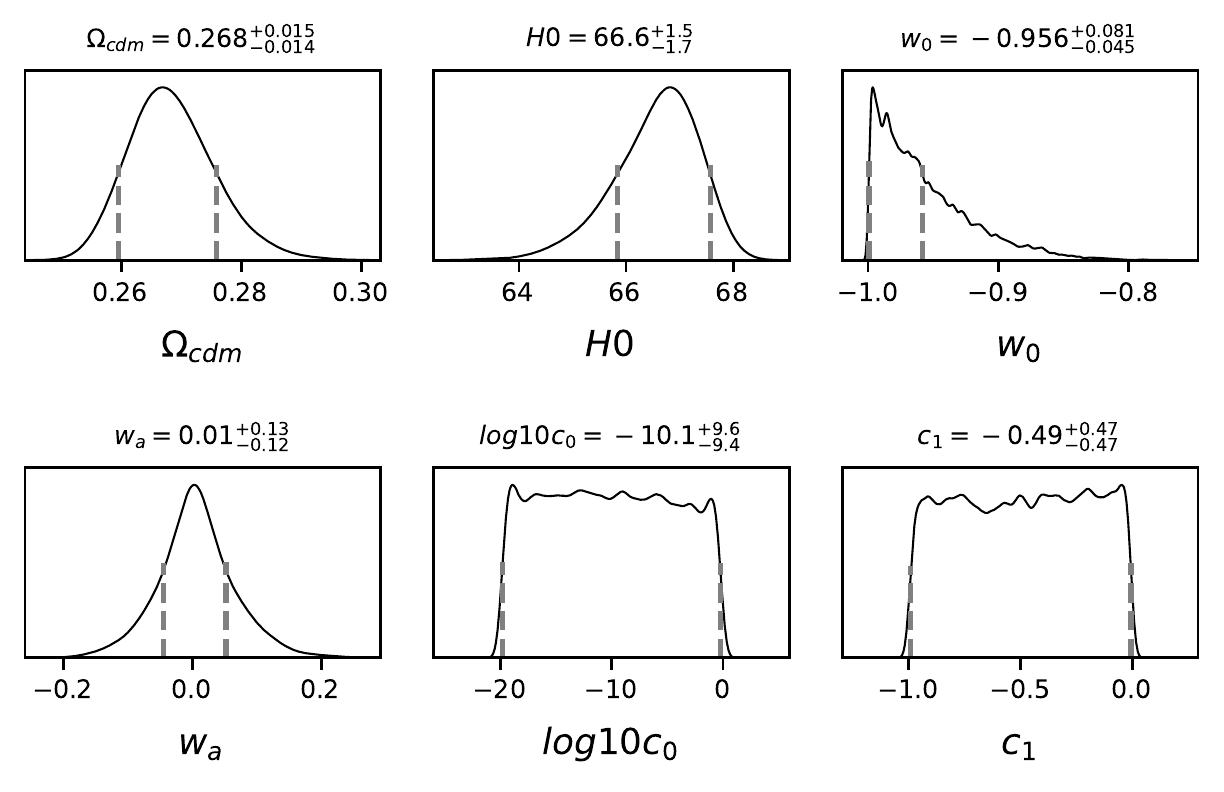}
    \caption{1-dimensional posterior distributions. Vertical dashed lines mark half-maximum x-coordinates while the limits quoted at top are $2\sigma$ limits.}
    \label{fig:1d_c_intervals}
\end{figure}

We adopt following parametric form for $c_s^2$: 
\begin{equation}
c_s^2 = c1*w + c0
\end{equation}
This is the simplest form that can capture both quintessence and
tachyonic models.
For quintessence, we have $c1 = 0$ and $c0=1$ and for tachyonic models
$c1=-1$ and $c0=0$.
We then do an MCMC sampling using CLASS with MontePython
\cite{Audren:2012wb,Brinckmann:2018cvx}.  
We use  CMB (Planck 2018 high-l TT,TE,EE, low-l EE, low-l TT,
lensing) \cite{2020A&A...641A...6P} and BAO data (Boss Data Release
12 \cite{2017MNRAS.470.2617A,2018JCAP...01..008B}, small-z BAO data
from 6dF Galaxy Survey \cite{2011MNRAS.416.3017B} and SDSS DR7 main
Galaxy sample \cite{2015MNRAS.449..835R}). 
We find that the two parameters $c1$ and $c0$ remain unconstrained.
In fig \ref{fig:traingle}, we show triangle plots for 2d marginalized
credible intervals.
Parameters relating to DE speed of sound are unconstrained.
This result is similar to analyses with constant $c_s^2$ have obtained
earlier 
\cite{PhysRevD.81.103513,2015MNRAS.452.2930M,2003MNRAS.346..987W,Bean:2003fb}.
While the previous work deals with either constant $c_s^2$ or some
particularly chosen form, here we have chosen an explicit
parameterized form for it, which encapsulates both quintessence and
tachyonic field.
In figure \ref{fig:1d_c_intervals}, we plot the marginalised posteriors. 

\section{CMB data and tachyonic models}

We modify CLASS to implement tachyonic models as a scalar field at
linear level, where equations are obtained from Lagrangian
corresponding to tachyonic dark energy (The equations and
modifications related information is provided in \ref{app A}).
Two potentials which we code in CLASS are: 
\begin{itemize}
\item 
  Exponential potential
  \begin{equation}
    V(\phi) = V_0 \exp(-\frac{\phi}{\phi_a})
  \end{equation}
\item
  Inverse Square Potential
  \begin{equation}
    V(\phi) = \frac{n}{4\pi G}\left(1-\frac{2}{3n}  \right)^{\frac{1}{2}}\frac{1}{\phi^{2}}
  \end{equation}
\end{itemize}
These two potentials have some interesting features and have been
studied in detail \cite{Padmanabhan:2002cp,Sen_2002}.
These potentials have been constrained using low red-shift data in
\cite{Singh_2019,Singh_2020}.
In \cite{Singh_2019}, tachyonic models were constrained using low
redshift data from supernova, Hubble parameter measurements, and BAO
data.
Evolution of perturbations was considered in \cite{Singh_2020} and
redshift space distortion data was used for model comparisons.
These models have not yet been constrained using CMB data.
We use CLASS with MontePython to constrain the tachyonic models (with
the above-mentioned potentials) using Planck 2018 data
\cite{2020A&A...641A...1P}. We use the following combinations of data:  
\begin{itemize}
\item 
  CMB (Planck 2018 high-l TT,TE,EE, low-l EE, low-l TT, lensing) \cite{2020A&A...641A...6P}
\item
  BAO (Boss Data Release 12 \cite{2017MNRAS.470.2617A,2018JCAP...01..008B}, small-z BAO data from 6dF Galaxy Survey \cite{2011MNRAS.416.3017B} and SDSS DR7 main Galaxy sample \cite{2015MNRAS.449..835R})
\item
  Combination of the above-mentioned CMB and BAO data.
\item
  JLA data \cite{2014A&A...568A..22B}.
\end{itemize}

\begin{figure}[hbt!]
    \centering
    \includegraphics[width=0.4\textwidth]{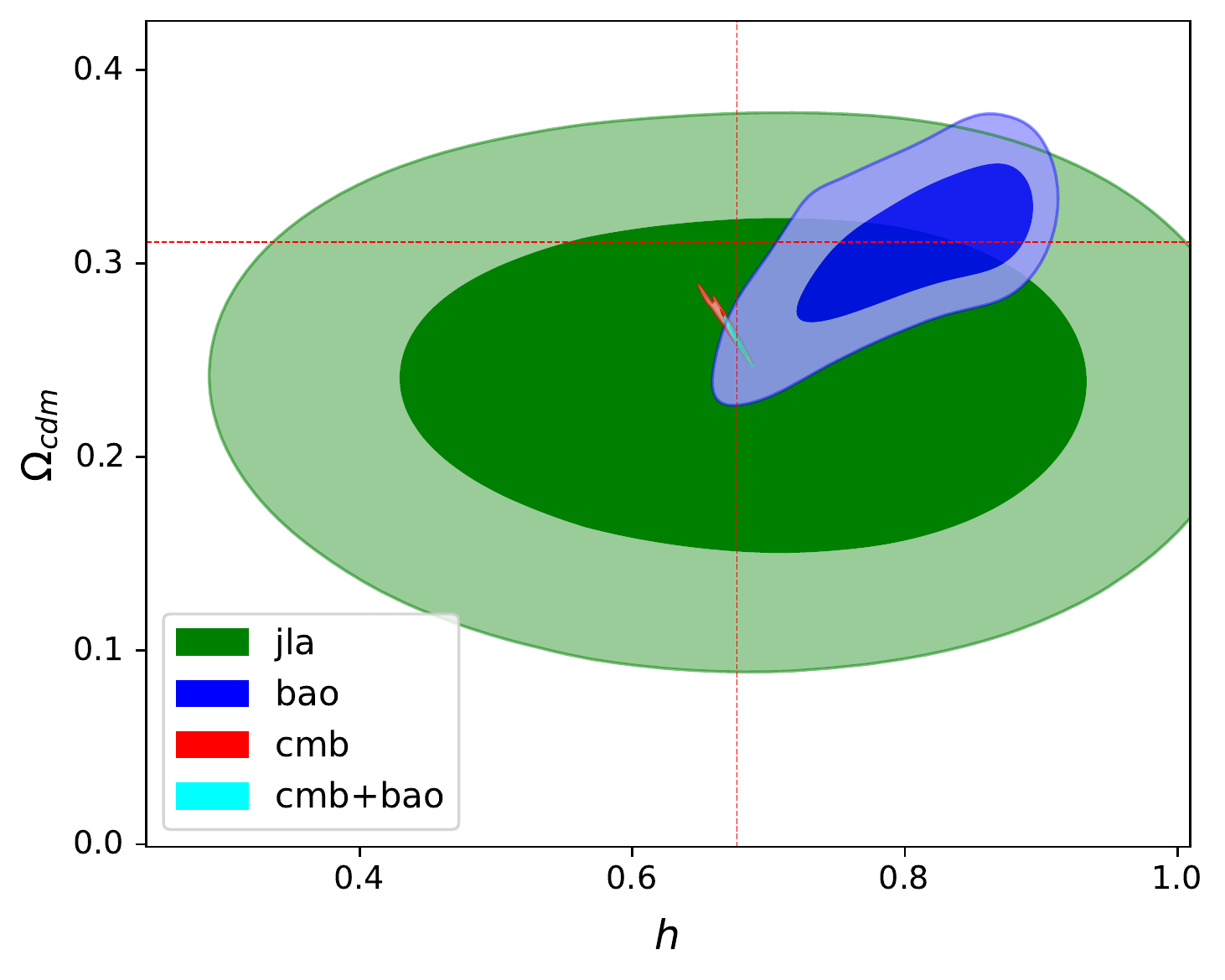}
    \includegraphics[width=0.4\textwidth]{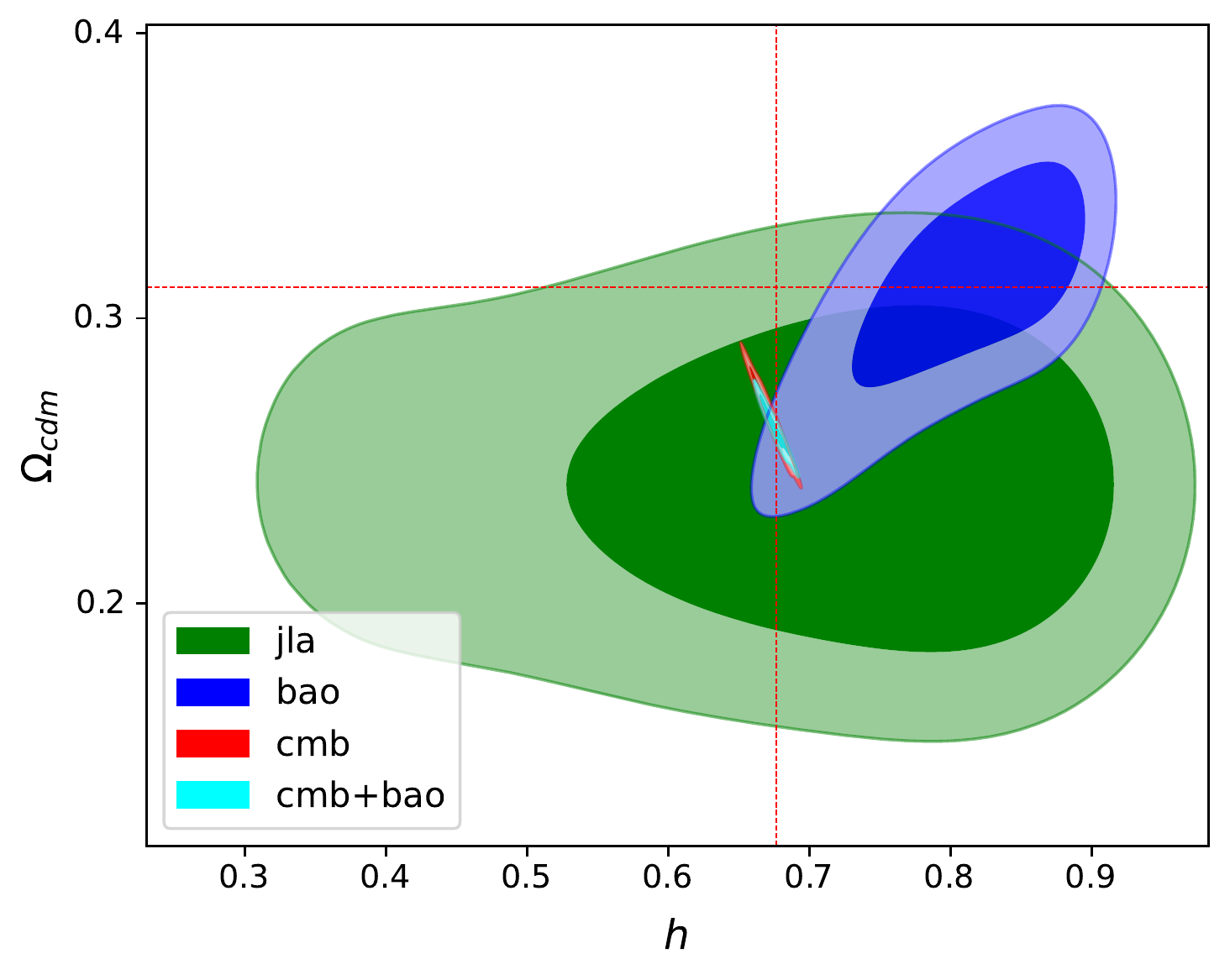}
    \caption{2-d plots for present day matter density parameter $\Omega_{cdm}$ and Hubble parameter ($h$). Left panel is for exponential potential while right one is for inverse square. Red lines show the best-fit values for $\Lambda CDM$ model from Planck 2018 \cite{2020A&A...641A...6P}. CMB data, as already known, shrinks the constrained region. $H_0$ tension is not resolved by tachyonic models considered here.}
    \label{fig:omega_h}
\end{figure}

\begin{figure}[hbt!]
    \centering
    \includegraphics[width=0.8\textwidth]{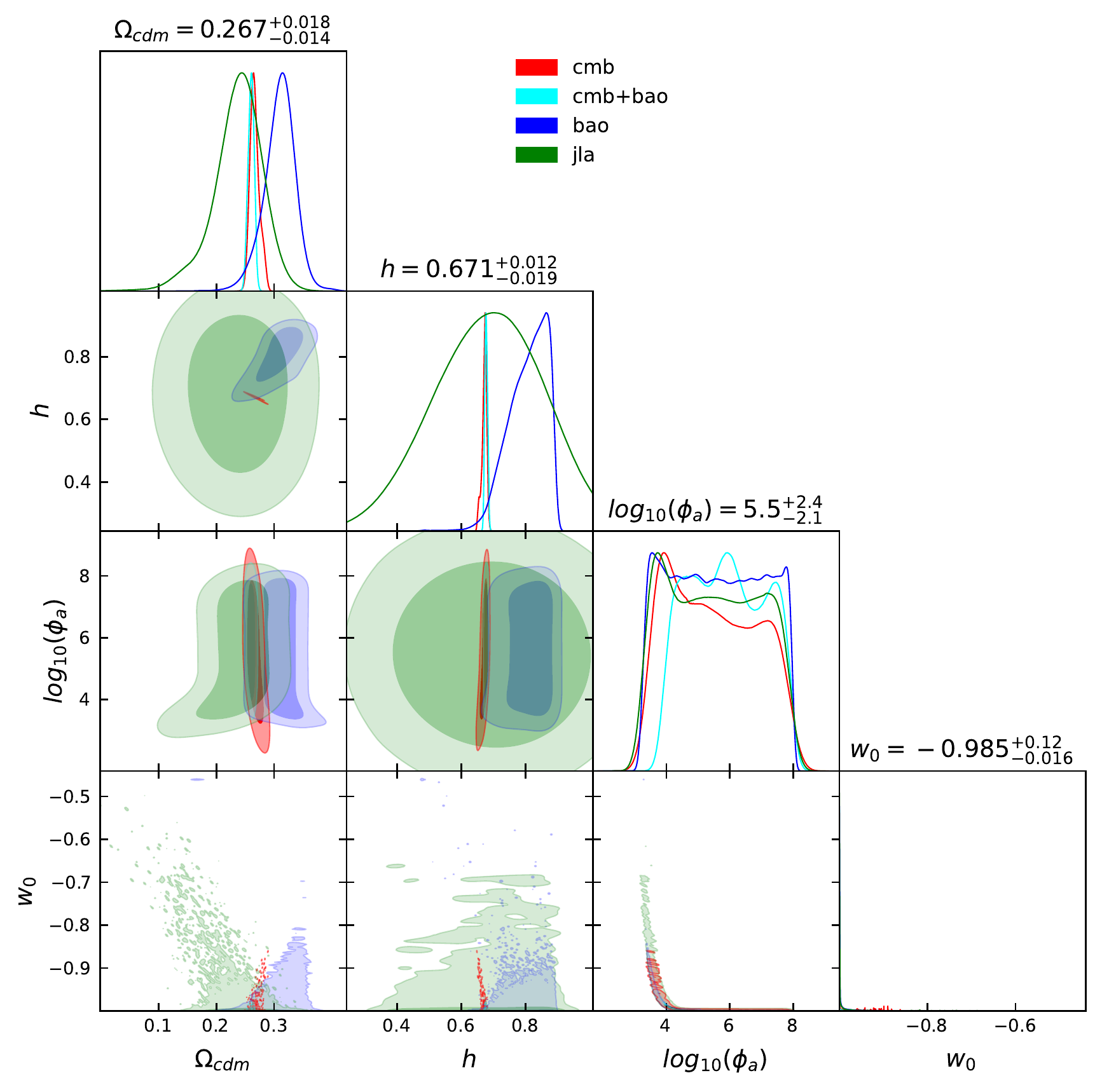}
    \caption{Triangle plot using four combinations of data, for exponential potential. Potential parameter $\phi_a$ is slightly constrained to be greater with certain minima. While $\phi_a$ appears to have nonlinear correlations with $w_0$, $w_0$ is constrained be close to $-1$. }
    \label{fig:exp_triangle}
\end{figure}

\begin{figure}[hbt!]
    \centering
    \includegraphics[width=0.8\textwidth]{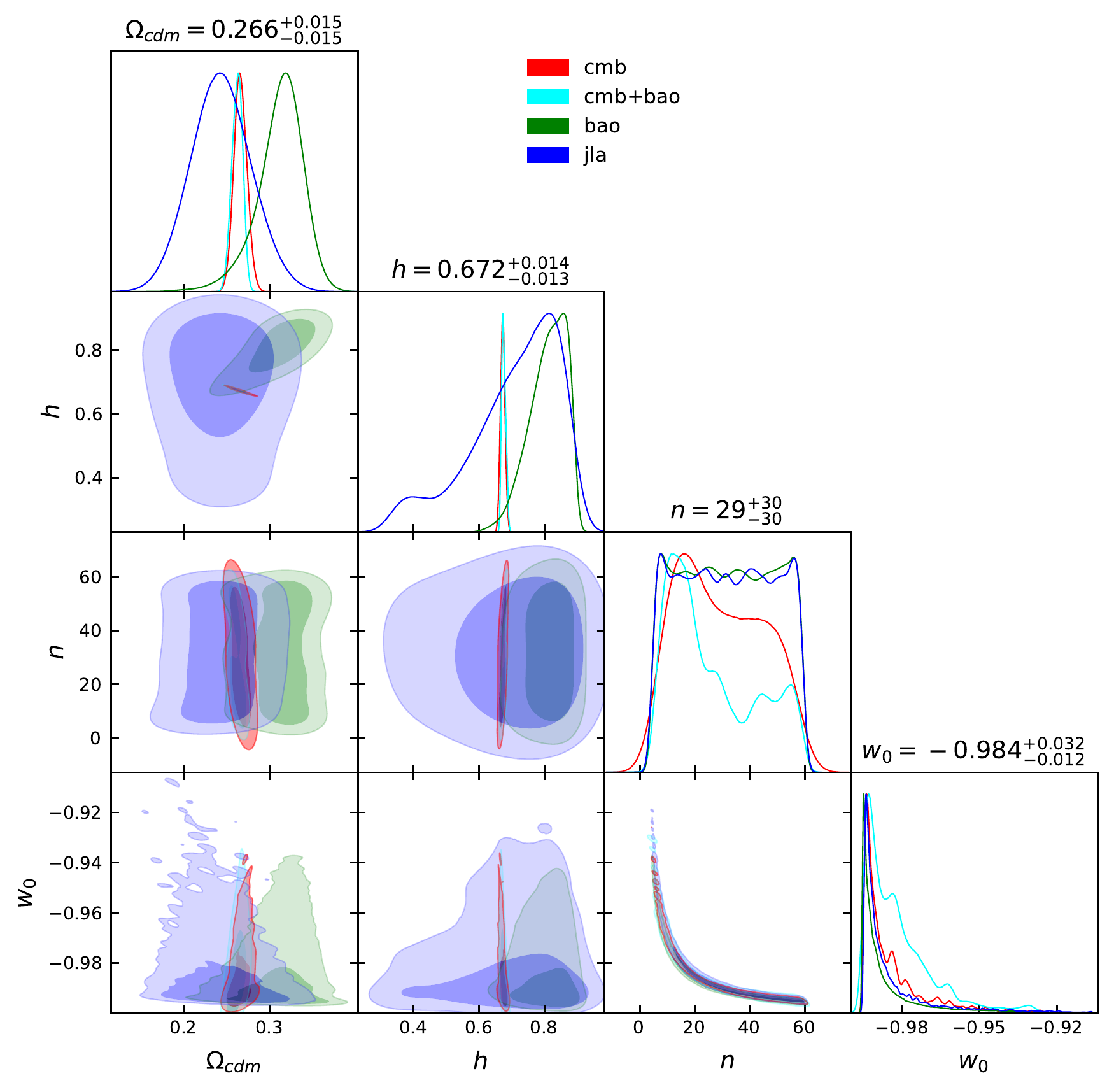}
    \caption{Triangle plot using four combinations of data, for inverse square potential. Results are somewhat similar to that for exponential case as potential parameter $n$ is slightly constrained and is correlated to $w_0$. A particular value of $w_0$ is favoured, which is not $-1$, but is close to it.}
    \label{fig:insq_triangle}
\end{figure}

\begin{figure}[hbt!]
    \centering
    \includegraphics[width=0.8\textwidth]{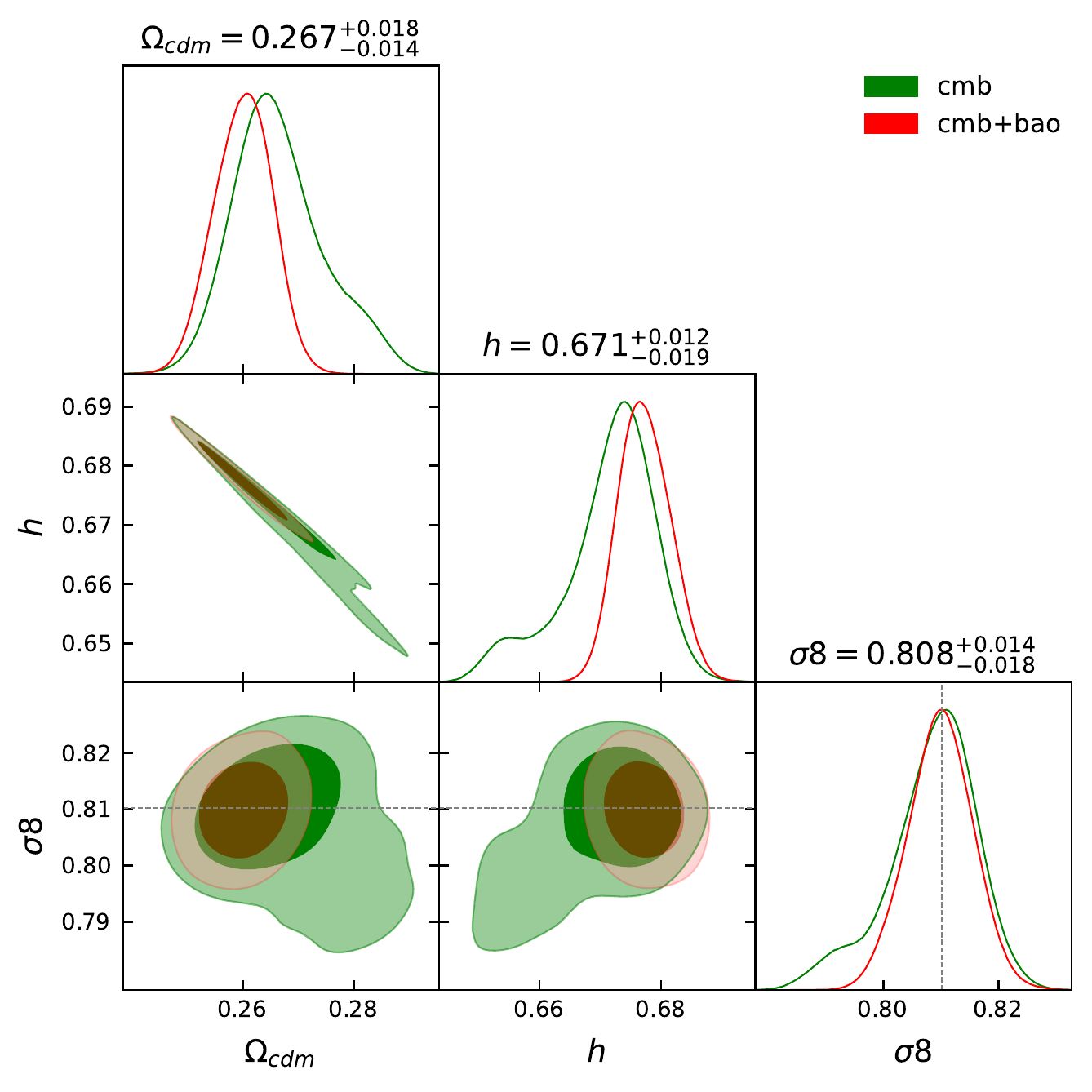}
    \caption{Triangle plot with $\sigma_8$ for the exponetial potential. The value of $\sigma_8$ is compatible with that for $\Lambda CDM$ from Planck 2018 (plotted as line in this figure).}
    \label{fig:exp_s8_triangle}
\end{figure}

\begin{figure}[hbt!]
    \centering
    \includegraphics[width=0.8\textwidth]{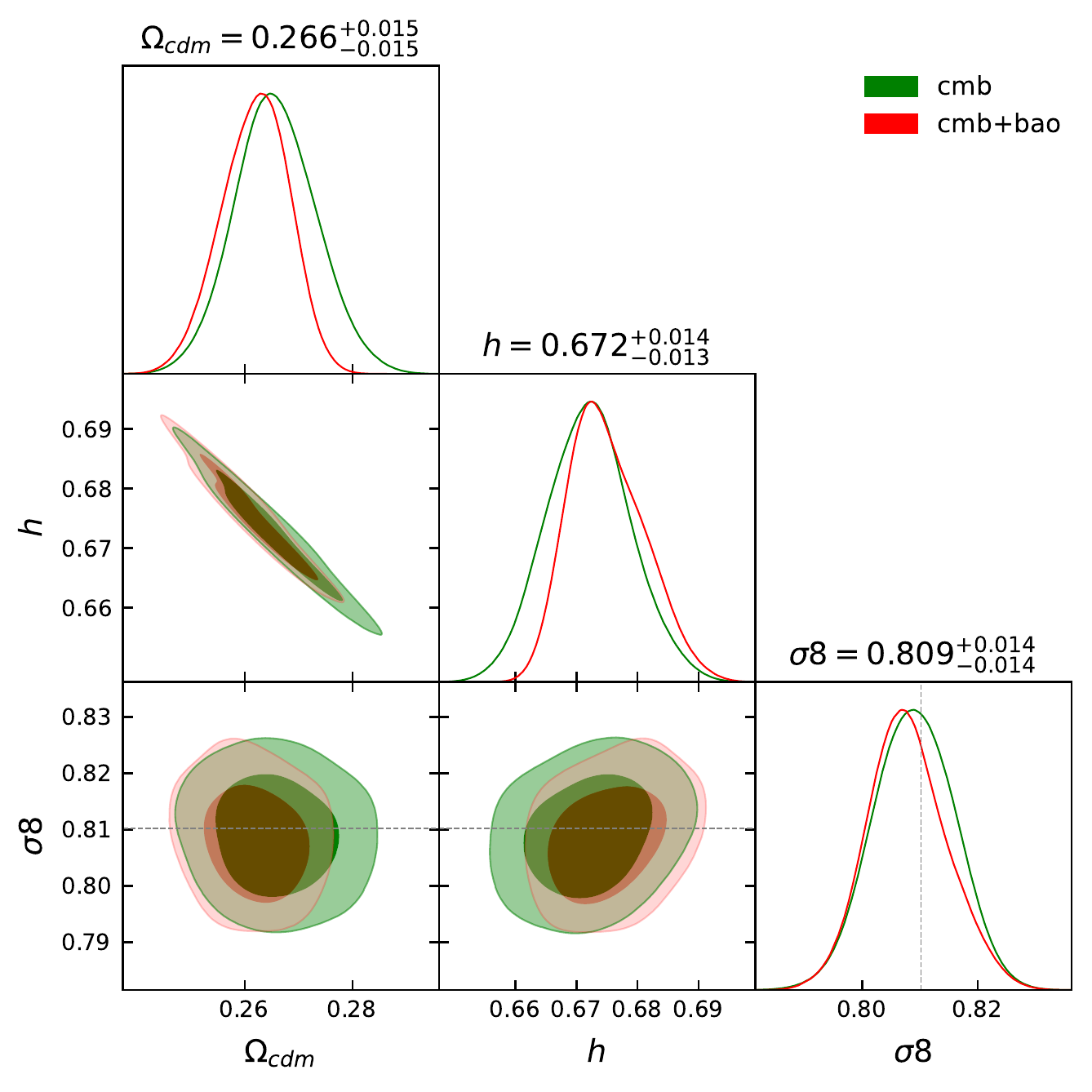}
    \caption{Triangle plot with $\sigma_8$ for the  inverse square potential. The constraints are comparable with those for exponential potential as well as $\Lambda CDM$ }
    \label{fig:insq_s8_triangle}
\end{figure}

\subsection{Results}

We first consider constraints on parameters that only concern
background evolution and are needed irrespective of potentials:
density parameter for dark matter and the Hubble constant.   
In figure \ref{fig:omega_h}, we plot contours for present-day matter
density contrast ($\Omega_{cdm}$) and dimensionless Hubble constant
($h$).
CMB data provides tight constraints.
The best fit values of these parameters, from the $\Lambda CDM$ model
based CMB constraint in Planck 2018 cosmo parameters paper, is
represented by red lines in the figure.
We find that the best-fit value (2-d) lies in the 1-sigma region of the JLA
data, but it is out of 2-sigma regions for CMB and BAO data
constraints.
While $h$ is consistent (within 2-sigma regions), it is
$\Omega_{cdm}$, which is lower for these tachyonic field based
cosmological models.
So, inference of dark matter content of the Universe shows dark energy
model dependence, when considering extensions beyond $\Lambda$.
In figure \ref{fig:exp_triangle}, we present the triangle plot for
exponential (exp) potential with potential parameter $\phi_a$ and
present-day equation of state $w_0$, included along with density and
Hubble parameter.
$w_0$ is constrained to be close to $-1$.
The potential parameter $\phi_a$ is not constrained by any of the used
data.
Triangle plot for inverse-square (insq) potential is presented in
figure \ref{fig:insq_triangle}.
Again, potential parameter $n$ is allowed a very wide region and $w_0$
is very close to $-1$.
Plots with $\sigma_8$ are shown in figures \ref{fig:exp_s8_triangle}
and \ref{fig:insq_s8_triangle}.
The constraints for $\sigma_8$ for two potentials agree with each
other as well as with that for $\Lambda CDM$.
This is again a manifestation of the fact that the models which have
same background evolution and are close to $\Lambda$ are extremely
difficult to distinguish.  

\section{Summary \& Prospects}

We have studied the prospects of using linear perturbation theory to
distinguish two different models of dark energy: quintessence and
tachyonic field.
Specifically, we investigate the differences in dynamics of
perturbations for the same background expansion in both models.
This helps us separate the effects coming from different background
expansions and differences due to perturbations. 

We recast linear theory equations in a form that provides insight into
how the systems of perturbations differ in two theories.
We show that when the equations for both are written in fluid terms,
substituting for corresponding field terms, one of the equations has
extra terms for quintessence.
These first-order terms are multiplied by $\omega\equiv (1+w)$.
This implies that if the background expansion is close to $w=-1$,
differences between the two models diminish.

We calculated and showed the evolution of quantities like $\psi$ and
its derivative, which affects the observables.
These numerical calculations demonstrate the theoretical dependence on
the factor of  $(1+w)$.

We find that the differences between models while being small at all
scales are largest around the scale of $10000$~Mpc.
We believe that this is due to the difference between the effective
speed of sound in two models and that this difference is seen in the
transition scales from suppression of perturbations at small scales to
growth at large scales. 

We used the definition of effective $c_s^2$ for two models to write a
parametric form for $c_s^2$($\equiv c1*w + c0$)  which incorporates
both fields as instances of particular values of the parameters.
We then used CMB data to constrain this parametric form to see if we can
distinguish two models and find that two parameters $c0$ and $c1$
remain unconstrained.

We modified the CMB anisotropy code CLASS to incorporate tachyon
models.
We then used it to constrain common tachyonic potentials:
$V\propto \exp(-\frac{\phi}{\phi_a})$ and $V\propto \phi^{-2}$ using
CMB and other data.
We find that the parameters are very weakly
constrained.

We have shown that it is very difficult to distinguish between these
two classes of models at large scales where linear perturbation theory
is applicable.   
We have also shown that this is primarily because only models with
$(1+w) \ll 1$ are allowed and in this regime, the differences between
the two classes of models are effectively of second order.   
Combined with our earlier work where we have explored these models at
small scales using spherical collapse, it appears that there are no
obvious observables available at present that may be used to
distinguish between these two classes of models if the expansion
history is the same. 
We can conclude that at least for these two classes of dark energy, as
also for a fluid model of dark energy, the choice of class of models
is irrelevant and calculation of observables may be done in any model.  
On one hand, this is a potential simplification of calculations, on
the other hand, it means that we cannot know which of the models is
the true model for dark energy. 

We are exploring the following to broaden the scope of our conclusions.
\begin{itemize}
\item
Working out forecasts to find out the sensitivity of observations
required to differentiate between the class of models.
This will allow us to see whether future observations can potentially
distinguish between the two classes of models.
A comparison with the capabilities of upcoming surveys is required to
ascertain whether the problem can be solved in the coming years. 
\item
Generalize the analysis to other classes of models to check whether
the feature of a prefix of $(1+w)$ in model-dependent terms is generic
or specific to the classes studied here.
We would like to be able to check at least some classes of models with
minimal coupling. 
\item
Going beyond linear theory to see if a higher-order calculation or a
general numerical relativity calculation at small scales can bring out
some features that are not accessible in the two limiting cases we
have used so far. 
\end{itemize}

\section*{Acknowledgment}
All the computational work was done on computing facilities provided by IISER Mohali. This research uses NASA's astrophysics data system (ADS) services.

\clearpage

\appendix

\section{Equations in Synchronous Gauge}
\label{app A}

Here we present the equations required for modification of CLASS for
tachyonic field.
Since synchronous gauge is the default gauge in CLASS, we write the
equations in this gauge.
Quintessence with some potentials are already implemented in CLASS,
one can simply follow the same structure for incorporating the
tachyonic models.
Here, we present equations for both quintessence and tachyonic field
because this helps on modifications comparing with quintessence
implementation. 

Note: In this section, we use conformal time and prime represents
derivative wrt to conformal time.

Field dynamics equation:
For tachyonic models
\begin{eqnarray}
  (\delta\phi)^{\prime\prime} &= & \left[1-\frac{{\phi^\prime}^2}{a^2} \right]
\left[ a^2(\delta\phi) \left\lbrace  \frac{({V,}_{\phi})^2}{V^2} -
  \frac{({V,}_{\phi\phi})^2}{V^2}  \right\rbrace   +
  \nabla^2(\delta\phi) - \frac{\phi^\prime h^\prime}{2}  \right] 
\nonumber\\ 
&& + (\delta\phi)^\prime \left[ -\frac{2a^\prime}{a} +
  9\frac{a^\prime}{a} \frac{{\phi^\prime}^2}{a^2} +
  2\frac{{V,}_{\phi}}{V^2}{\phi^\prime}^2 \right] 
\end{eqnarray}
For quintessence
\begin{equation}
(\delta\phi)^{\prime\prime} = -a^2(\delta\phi)({V,}_{\phi\phi}) -
  \frac{\phi^\prime h^\prime}{2}
  -\frac{2a^\prime}{a}(\delta\phi)^\prime   + \nabla^2(\delta\phi)  
\end{equation}
Density ($\delta \rho$) and pressure perturbations ($\delta p$): 
For tachyonic field
\begin{equation}
  (\delta\rho) =
  \frac{(\delta\phi)({V,}_{\phi})}{\sqrt{1-\frac{{\phi^\prime}^2}{a^2}}}
  + \frac{V\phi^\prime (\delta\phi)^\prime}{a^2\left[
      1-\frac{{\phi^\prime}^2}{a^2} \right]^{3/2} } 
\end{equation}
\begin{equation}
  (\delta p) =
  -(\delta\phi)({V,}_{\phi})\sqrt{1-\frac{{\phi^\prime}^2}{a^2}} + 
\frac{V\phi^\prime (\delta\phi)^\prime}{a^2\sqrt{
    1-\frac{{\phi^\prime}^2}{a^2} } } 
\end{equation} 

For quintessence
\begin{equation}
(\delta\rho) = (\delta\phi)({V,}_{\phi}) + \frac{\phi^\prime (\delta\phi)^\prime }{a^2}
\end{equation}
\begin{equation}
(\delta p) = -(\delta\phi)({V,}_{\phi}) + \frac{\phi^\prime (\delta\phi)^\prime }{a^2}
\end{equation}
Effective velocity perturbations: \\
For tachyonic field
\begin{equation}
(\bar{\rho}+\bar{p}) \theta = i k^j(\delta T)^0_j = \frac{\phi^\prime}{a^2}k^2(\delta\phi) \frac{V}{\sqrt{1-\frac{{\phi^\prime}^2}{a^2} }}
\end{equation}
For quintessence
\begin{equation}
(\bar{\rho}+\bar{p}) \theta =  \frac{\phi^\prime}{a^2}k^2(\delta\phi)
\end{equation}

\section*{References}
\bibliography{ref}

\bibliographystyle{ieeetr}

\end{document}